\documentclass{iaus}
\usepackage{graphicx}

\title[Lithium in the Globular Cluster NGC 6397] 
{Main-Sequence and sub-giant stars in the Globular Cluster NGC6397:
The complex evolution of the lithium abundance} 

\author[Gonz\'alez Hern\'andez et al.]   
{J. I. Gonz\'alez Hern\'andez$^{1,2}$\thanks{Present address: Dpto. de Astrof\'{\i}sica y Ciencias de la Atm\'osfera, Facultad de
F\'{\i}sica, Universidad Complutense de Madrid, E-28040 Madrid,
Spain. Email: {\tt jonay@astrax.fis.ucm.es}}, P. Bonifacio$^{1,2,3}$,
E. Caffau$^1$, M. Steffen$^4$, \\  H.-G. Ludwig$^{1,2}$, N.
Behara$^{1,2}$, L. Sbordone$^{1,2}$, R. Cayrel$^1$, \and S.~Zaggia$^5$
}

\affiliation{
$^1$GEPI, Observatoire de Paris, CNRS, Universit\'e Paris Diderot; \\[\affilskip]Place
Jules Janssen 92190 Meudon, France \\ email: {\tt
Jonay.Gonzalez-Hernandez@obspm.fr}\\[\affilskip] 
$^2$Cosmological Impact of the First STars (CIFIST) Marie Curie Excellence
Team\\[\affilskip]
$^3$Istituto Nazionale di Astrofisica - Observatorio\\[\affilskip] 
Astronomico di Trieste, Italy\\[\affilskip]
$^4$Astrophysikalisches Institut Potsdam, An der Sternwarte 16, \\[\affilskip] 
D-14482 Potsdam, Germany \\[\affilskip]
$^5$INAF - Osservatorio Astronomico di Padova, \\[\affilskip] 
Vicolo dell'Osservatorio 5, Padua 35122, Italy\\[\affilskip] 
}

\pubyear{2009}
\volume{268}  
\pagerange{xxx--xxx}
\date{15 December 2009 and in revised form 15 December 2009}
\jname{Light Elements in the Universe}
\editors{Corinne Charbonnel, Monica Tosi, Francesca Primas\\ \& Cristina Chiappini, eds.}
\begin{document}

\maketitle

\begin{abstract}

Thanks to the high multiplex and efficiency of Giraffe at
the VLT we have been able for the first time to observe the Li I
doublet in the Main Sequence stars of a Globular Cluster. At the same
time we observed Li in a sample of Sub-Giant stars of the same B-V
colour. 

Our final sample is composed of 84 SG stars and 79 MS stars.
In spite of the fact that SG and MS span the same temperature range we
find that the equivalent widths of the Li I doublet in SG stars are 
systematically larger than those in MS stars, suggesting a higher Li
content among SG stars. This is confirmed by our quantitative analysis
carried out making use of 1D hydrostatic plane-parallel models and 3D
hydrodynamical simulations of the stellar atmospheres. 

We derived the effective temperatures of stars in our the sample from
H$\alpha$ fitting. Theoretical profiles were computed using 3D
hydrodynamical simulations and 1D ATLAS models. Therefore, we are able
to determined 1D and 3D-based effective temperatures. We then infer Li
abundances taking into account non-local thermodynamical equilibrium
effects when using both 1D and 3D models.

We find that SG stars have a mean Li abundance higher by
0.1\,dex than MS stars. This result is obtained using both 1D and 3D
models. We also detect a positive slope of Li abundance with effective
temperature, the higher the temperature the higher the Li abundance,
both for SG and MS stars, although the slope is slightly steeper for
MS stars. These results provide an unambiguous evidence that the Li
abundance changes with evolutionary status.  

The physical mechanisms responsible for this behaviour are not yet
clear, and none of the existing models seems to describe accurately 
these observations. Based on these conclusions, we believe that the 
cosmological lithium problem still remains an open question. 

\end{abstract}

\firstsection 
\section{Introduction}

The determination of the baryonic density from the fluctuations of
the cosmic microwave background (CMB) by the WMAP satellite
(\cite[Spergel et al. 2007, Cyburt et al. 2008]{spe07,cyburt}) implies
a primordial Li abundance which is $\log ({\rm Li}/{\rm H})+ 12
=2.72\pm0.06$, at least 0.3--0.5~dex higher than the Li abundance
determined in metal-poor stars of the Galactic halo (\cite[Spite \&
Spite 1982]{sas82}).

Many different models of Li depletion have been proposed to explain
discrepancy:  
(a) \cite{pia06} proposed that the first generation
of stars, Population III stars, could have processed some fraction of
the halo gas, lowering the lithium abundance;
(b) other authors suggest that the primordial Li abundance has been
uniformly depleted in the atmospheres of metal-poor dwarfs by some
physical mechanism (e.g. turbulent diffusion as in \cite{ric05,kor06};
gravitational waves as in \cite{cat05}, etc.); and (c) finally, it has
been also suggested that the standard Big Bang nucleosynthesis (SBBN)
calculations should be revised, possibly with the introduction of new
physics as in e.g. \cite{jed04,jed06,jittoh,hisano}. 

Here we present the determination of Li abundances of subgiant (SG) 
and main-sequence (MS) stars of the cluster NGC~6397. 
This work provides the first observations of the Li doublet in
MS stars of a Globular Cluster. 

\section{Observations}

We performed spectroscopic observations of the globular cluster 
NGC~6397 with the multi-object spectrograph FLAMES-GIRAFFE at
the VLT on 2007 April, May, June and July, covering the spectral
range $\lambda\lambda$6400--6800 {\AA} at resolving power
$\lambda/\delta\lambda\sim17,000$.  

We selected subgiant and dwarf stars in the colour range
$B-V=0.60\pm0.03$, which ensures that both set of stars fall in a
similar and narrow effective temperature range (see Fig.~3 online in
\cite[Gonz\'alez Hern\'andez et al. 2009a]{gon09a}).

The spectra were reduced with the ESO pipeline and later on treated
within MIDAS. We correct the spectra for sky lines and, barycentric
and radial velocity. We typically combine 17 spectra of dwarfs and 4
spectra of subgiants to achieve a similar S/N ratio between 80 and 130
in both sets of stars. The mean radial velocity of the cluster stars
is $v_r=18.5$~km~s$^{-1}$. 

\begin{figure}
\centering
\includegraphics[height=6.7cm,angle=90]{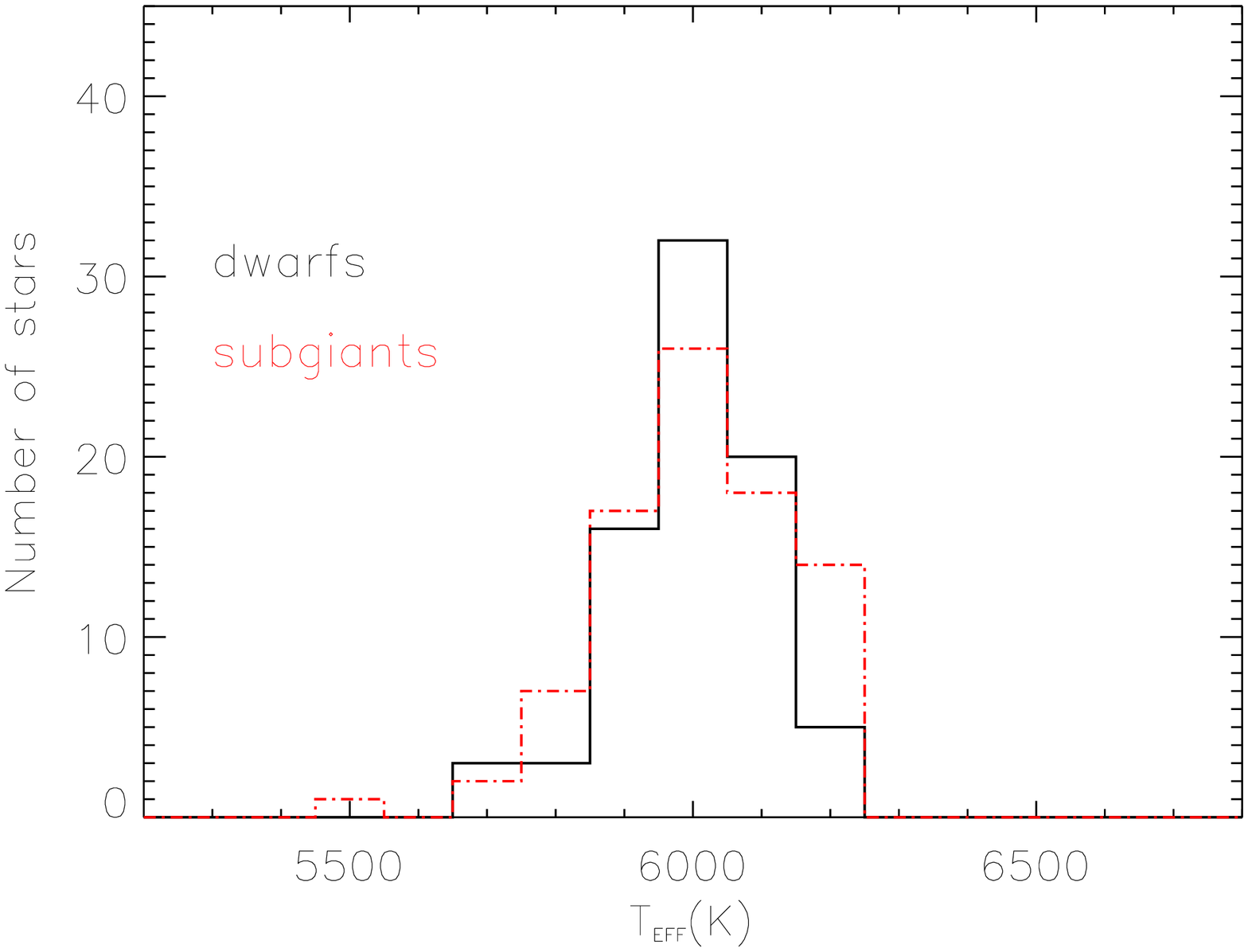}
\includegraphics[height=6.7cm,angle=90]{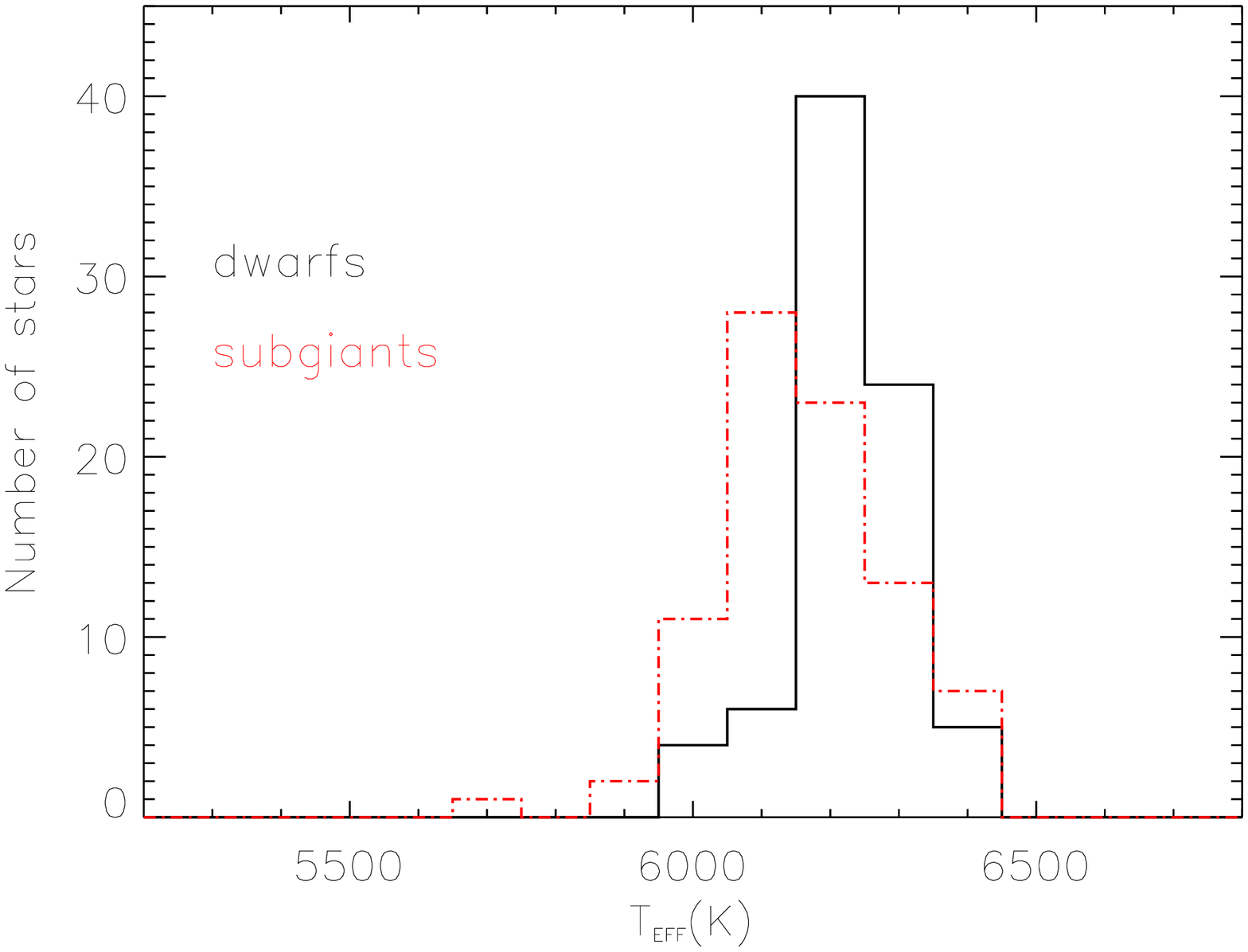}
\caption{\footnotesize{Histograms of 1D (left panel) and 3D (right
panel) effective temperatures derived by fitting the observed
H$\alpha$ profiles with theoretical profiles 
computed using 1D (left panel) and 3D (right panel) model atmospheres,
in bins of 100~K, in MS (solid line) and SG (dashed-dotted line) stars
in the Globular cluster NGC 6397.}}
\label{figteff} 
\end{figure}

\section{Stellar parameters}

We derived the effective temperature by fitting the observed
H$\alpha$ line profiles with synthetic profiles, using 3D
hydrodynamical model atmospheres computed with the CO$^5$BOLD code.
The details of this code are provided in \cite{fre02} and \cite{wed04}.  
The ability of 3D models to reproduce Balmer line profiles has been
shown in \cite{beh09}. In that work, the H$\alpha$ profiles of the Sun
as well as the metal-poor stars HD 84937, HD 74000 and HD 140283 are
studied. \cite{lud09} also quantified, from a purely theoretical
point of view, the difference between the effective temperatures derived
from H$\alpha$ fitting using 1D and 3D models.

We also derived the effective temperatures of MS and SG stars 
using 1D ATLAS~9 model atmospheres (see \cite[Kurucz 2005]{kur05})
and the same fitting procedure. In Fig.~\ref{figteff} we display the
histograms of the effective temperatures derived for MS and SG stars
using both 1D and 3D models. In the 1D case we got similar effective
temperatures for both sets of stars. However, using 3D hydrodynamical
models we obtained hotter temperatures by approximately 250~K in MS
stars and 150~K for SG stars. In the 3D case the histogram
of SG stars is slightly shifted with respect to the histogram of MS
stars to cooler temperatures, as expected
from the difference in surface gravity between MS and SG stars and the
sensitivity of the $B-V$ colour to the surface gravity.

Fixed values for the surface gravity were adopted for both subgiant
and dwarf stars in the sample, according to the values that best match
the position of the stars on a 12 Gyr isochrone (\cite[Straniero et
al. 1997]{str97}). The adopted values were $\log (g/{\rm
cm~s}^2)=4.40$ and 3.85 for MS and SG stars, respectively. 

\section{Li abundances}

We measure the equivalent width (EW) of the Li~{\scriptsize
I}~6708~{\AA} line in SG and MS stars by fitting synthetic spectra of
known EW to the observed Li profiles. 
\cite{gon09a} showed the histograms of the EWs
measured in SG and MS stars of this cluster (see their Fig.~1). In
that figure it is clearly seen that the EWs of SG stars are larger
than those of MS stars.
They also estimate the weighted mean EW of the SG stars, being $\sim
1.1$~pm larger than the weighted mean EW of MS stars. 
Although the colour $B-V$ is sensitive to surface gravity, a priori, 
this result was not expected, and clearly suggests that subgiants in
this cluster have actually higher Li abundances than dwarfs. 

We derived Li abundances using 3D model atmospheres. The line
formation of Li was treated in non-local thermodynamical equilibrium
(NLTE) using the same code and model atom used in \cite{cay07}. 
The analysis was also done using 1D model
atmospheres providing essentially the same picture, even when 
$T_{\rm eff}$ in 1D are lower (see also Fig.~6 online in
\cite[Gonz\'alez Hern\'andez et al. 2009a]{gon09a}).
In the 1D case we used the \cite{carlsson} NLTE corrections.

\begin{figure}
\centering
\includegraphics[height=6.7cm,angle=90]{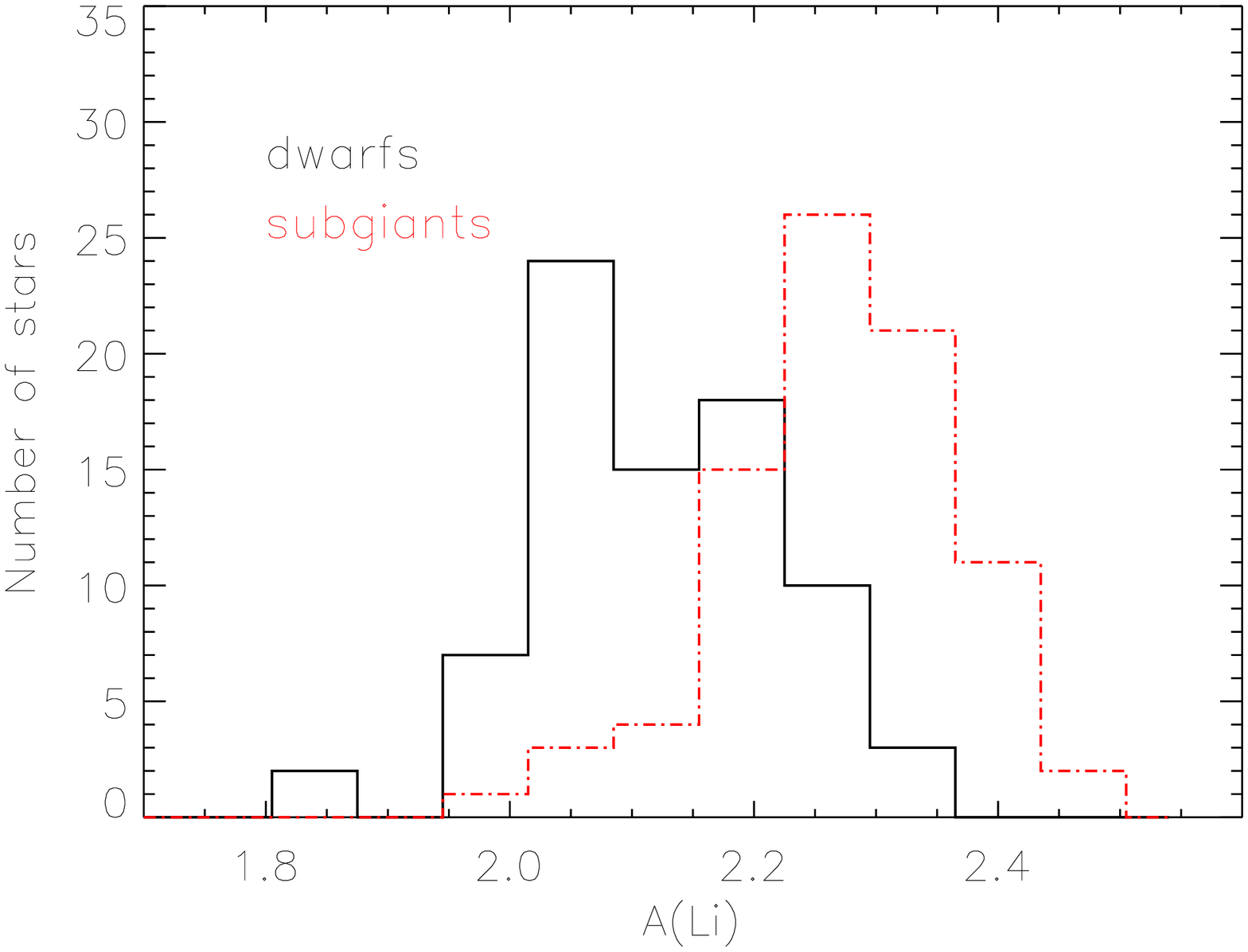}
\includegraphics[height=6.7cm,angle=90]{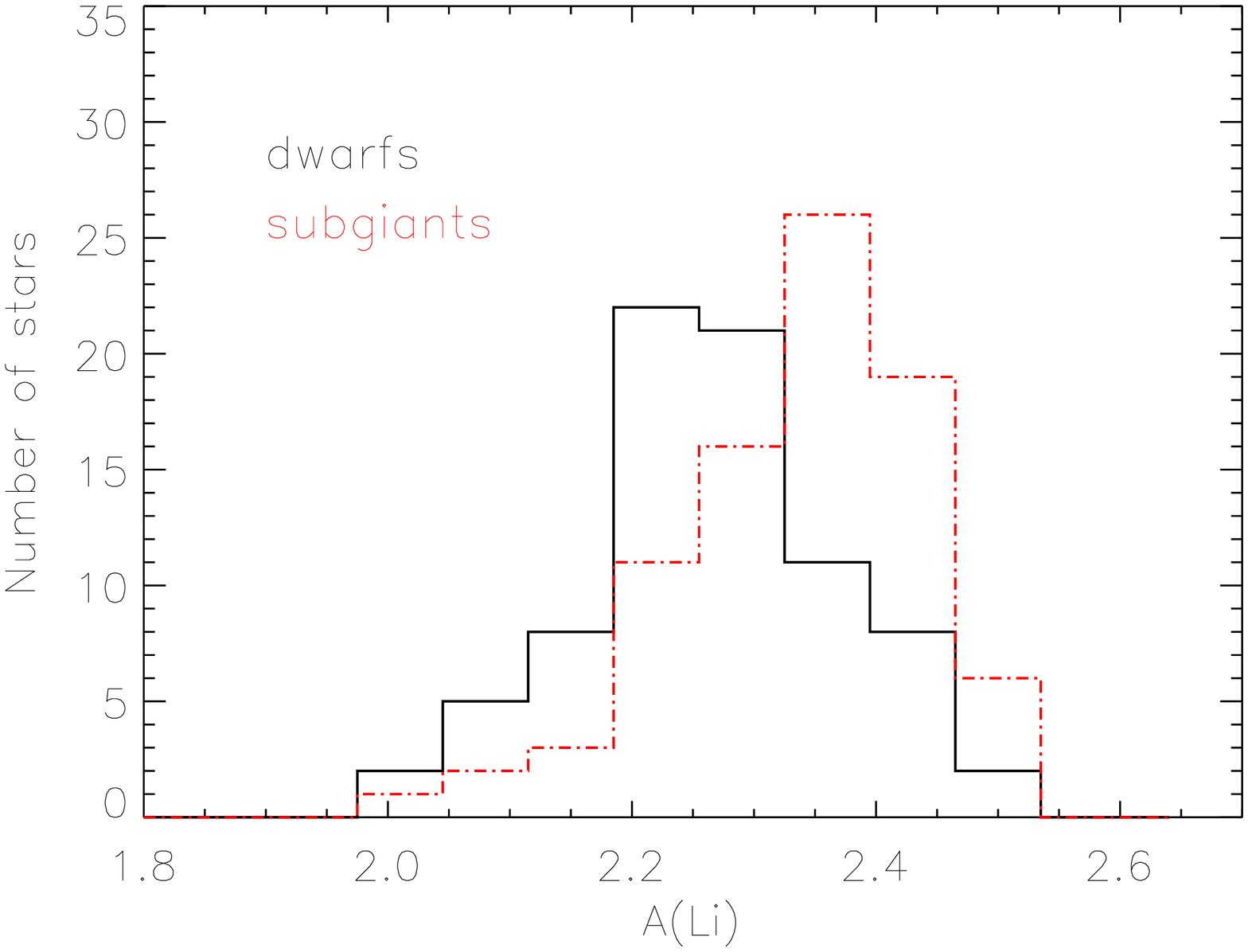}
\caption{\footnotesize{Histograms of 1D (left panel) and 3D (right
panel) non-LTE Li abundances, in bins of 0.07~dex, in MS (solid line)
and SG (dashed-dotted line) stars in the Globular cluster NGC 6397}}
\label{figali} 
\end{figure}

\section{Discussion and Conclusions}

In Fig.~\ref{figali} we display the histograms of the derived 1D-NLTE
and 3D-NLTE Li abundances of dwarf and subgiant stars of the globular
cluster NGC 6397. In both 1D and 3D cases, the SG stars have on
average larger amounts of Li content in their atmospheres than MS
stars.
The difference in the mean Li abundance of dwarfs and subgiants is
$\sim0.14$~dex in the 1D case and $\sim0.07$~dex in the 3D case. This
difference between the 1D and the 3D cases
is due to the effective temperatures derived using
1D and 3D models, and is not related to the NLTE corrections which
were independently computed in the 1D and 3D cases. In fact, due to
the cooler temperatures derived in 1D with respect to the 3D case, the
mean 1D Li abundance lower by $\sim0.13$~dex for the MS stars 
 $\sim0.06$~dex for the SG stars than the mean 3D abundance.

\cite{lin09} also find different mean Li abundances, using 1D models,
in MSs and SGs, but only by 0.03~dex although still significant at
1$\sigma$. 
However, their result is partially affected by the very narrow range 
of $T_{\rm eff}$ for MSs deduced by \cite{lin09} ($\sim 80$~K)
compared to the wide range  ($\sim 450$~K) for the SGs (see Fig.~7
online in \cite[Gonz\'alez Hern\'andez et al. 2009a]{gon09a}).

\cite{gon09a} showed, in their Fig.~2, the 3D-NLTE abundances of SG
and MS stars versus 3D effective temperatures (see Fig.~1 in
\cite{gon09b} for a similar picture but with $T_{\rm eff}$ and Li
abundances computed using 1D models).
The points in that figure display a decreasing trend of Li abundance
with decreasing temperature.
This lithium abundance pattern is different from what is
found among field stars (see e.g. \cite[Mel\'endez \& Ram{\'\i}rez
2004]{mel04}, \cite[Bonifacio et al. 2007]{bon07}, \cite[Gonz\'alez
Hern\'andez et al. 2008]{gon08}).  

Our results imply that the Li surface abundance depends on the
evolutionary status of the star.
In Fig.~2 of \cite{gon09a}, the Li isochrones for
different turbulent diffusion models (\cite[Richard et al.
2005]{ric05}). These models were
shifted up by 0.14 dex in Li abundance to make the initial
abundance of the models, $\log ({\rm Li}/{\rm H})=2.58$, coincide
with the primordial Li abundance predicted from fluctuations of the
microwave background measured by the WMAP satellite (\cite[Cyburt et
al. 2008]{cyburt}). 
The models assuming pure atomic diffusion, and, among those including
low levels of turbulent mixing, in particular the model T6.0,
preferred model in \cite{kor06} and \cite{lin09} are ruled out by our
observations, since they predict the Li content in MS stars to be
higher than in SG stars. 
The only model that predicts a Li pattern which is qualitatively
similar to that observed, is the T6.25 model. This model shows a
decreasing trend of Li abundances with decreasing temperatures and
also predicts higher abundances for SG stars than for MS stars
although at temperatures lower than 6000~K, which is not consistent
with the observed trend even in the 1D case (see Fig.~1 in
\cite[Gonz\'alez Hern\'andez et al. 2009b]{gon09b}).

Models including atomic diffusion and tachocline 
mixing (\cite[Piau 2008]{pia08}) do not seem to reproduce our 
observations either, since they provide a constant Li abundance up to
5500\,K. 
More sophisticated models are required, for instance, 
those models that besides including diffusion and rotation also
take into account the effect of internal gravity waves (\cite[Talon \&
Charbonnel 2004]{tac04}), seem to predict accurately the Li abundance pattern in
solar-type stars, at solar metallicity (\cite[Charbonnel \& Talon
2005]{cat05}), but models at low metallicity are still needed. 

The cosmological lithium discrepancy still needs to be solved.
Given that none of the existing models of Li evolution
in stellar atmospheres matches our observations in the globular
cluster NGC~6397, we hope
our results will prompt further new theoretical investigations.

\end{document}